\documentclass[]{spie}  

\usepackage{amsmath,amsfonts,amssymb}
\usepackage{graphicx}
\usepackage[colorlinks=true, allcolors=blue]{hyperref}

\usepackage{wrapfig}

\title{Generative Super-Resolution PET Imaging \\with Fourier Diffusion Models 
}

\author[a]{Matthew Tivnan}
\author[b]{Quanzheng Li
}
\affil[a]{Massachusetts General Hospital and Harvard Medical School, 55 Fruit St., Boston, MA, USA
}

\begin{document} 
\maketitle


\begin{abstract}
Neurological Positron Emission Tomography (PET) is a critical imaging modality for diagnosing and studying neurodegenerative diseases like Alzheimer’s disease. However, the inherent low spatial resolution of PET images poses significant challenges in clinical settings. This work introduces a novel Generative Super-Resolution (GSR) approach using Fourier Diffusion Models (FDMs) to enhance the spatial resolution of PET images. Unlike traditional methods, FDMs leverage the time-dependent Modulation Transfer Function (MTF) and Noise Power Spectrum (NPS) to generate high-resolution, low-noise images from low-resolution inputs. Our method was evaluated using simulated data derived from High-Resolution Research Tomograph (HRRT) PET images with 2 mm resolution. The results demonstrate that FDMs significantly outperform existing techniques, including  conditional diffusion models and image-to-image Schr\"{o}dinger bridge, across several metrics, including structural similarity and noise suppression. Our simulation results highlight the potential of FDMs to generate high-quality 2mm resolution reconstructions given 4mm resolution input PET data.
\end{abstract}



\section{INTRODUCTION}

Positron Emission Tomography (PET) imaging is a powerful tool for the diagnosis of neurodegenerative diseases such as Alzheimer's disease. The radiotracer fluorodeoxyglucose (FDG) is particularly significant because it acts as an analog of glucose, the primary energy source for the brain. FDG PET imaging allows for the visualization of metabolic activity in the brain, which is crucial for detecting abnormalities associated with neurodegenerative conditions \cite{dave2020fdg}. For instance, in Alzheimer’s disease, regions of decreased FDG uptake correspond to areas of neurodegeneration, aiding in both diagnosis and monitoring of disease progression \cite{rice2017diagnostic}. Beyond its application in neurology, PET imaging is also widely used in oncology for tumor detection, staging, and monitoring treatment response, as well as in cardiology for assessing myocardial viability, and in infection and inflammation imaging \cite{rohren2004clinical}.

While PET has high sensitivity to low concentrations of radiotracers, it also has low spatial resolution, presenting a significant challenge in accurately diagnosing and studying neurological diseases \cite{gong2016assessment}. The most widely available PET scanners are full body scanners with a spatial resolution ranging of approximately 4 mm full-width at half-maximum (FWHM), which is significantly lower than MRI and CT, which have less than 1 mm resolution \cite{vandenberghe2020state}. As a result of this low spatial resolution, PET images often fail to capture fine details, leading to less precise diagnoses and a poorer understanding of disease mechanisms. 

To address this issue, there has been significant research in the area of improving PET scanner instrumentation, through methods such as incorporating time-of-flight (TOF) technology or depth of interaction (DOI) measurements, which can help to enhance spatial resolution \cite{lewellen1998time, miyaoka1998design}. There are also dedicated neurological PET scanners with higher resolution than typical  resolution, such as the High Resolution Research Tomograph (HRRT), which achieves a spatial resolution of approximately 2 mm FWHM \cite{schmand1998performance, blinder2012scanning}. Recently, the NeuroExplorer PET (NX-PET) has achieved groundbreaking 1mm FWHM spatial resolution \cite{carson2023exceptional}. While these advanced scanners may be available at some research institutions, they are expensive and unlikely to be widely adopted for routine clinical scans. Therefore, there is an opportunity to use data collected from these advanced high resolution scanners to learn patterns of high spatial frequency distributions of radiotracers. By leveraging this data, we can train deep-learning-based probabilistic priors, such as score based diffusion models, to improve the quality of images produced by more widely available full body PET scanners with lower resolution. We refer to the task of estimating high spatial frequency radiotracer distributions given low resolution measurements as Generative Super-Resolution (GSR) PET Imaging. In this work, we investigate the potential of this approach to enhance the image quality of widely available PET systems. We believe that GSR has the potential to provide significant and widespread improvements to neurological PET for diagnosing neurodegenerative diseases.

\begin{figure}
    \centering
    \includegraphics[width=0.99\linewidth]{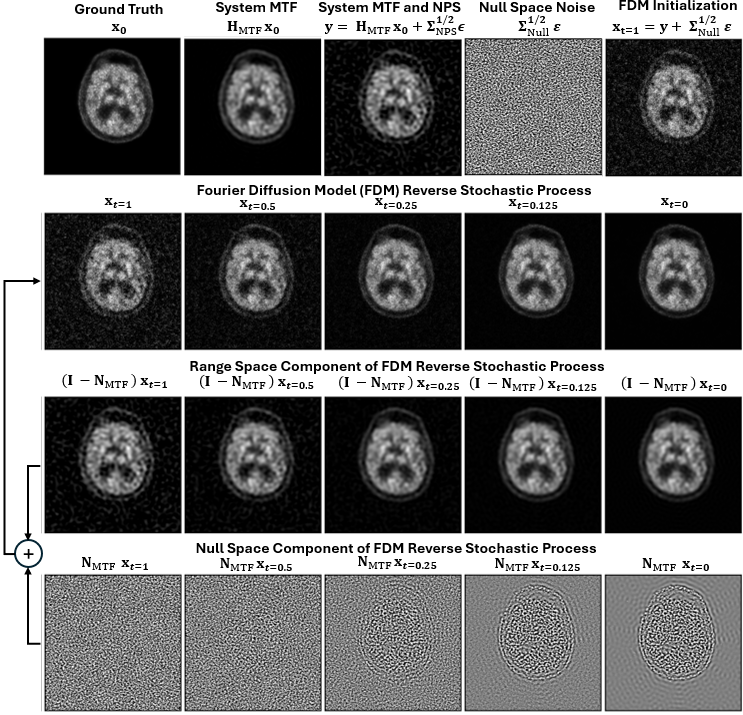}
    \caption{The top row shows the system MTF and NPS for the low resolution PET system. The FDM reverse process can be initialized by the measurements plus null space noise. The second row shows the FDM reverse diffusion process which samples from the posterior of high resolution PET images given low resolution PET images. The third and fourth row show a decomposition into range space and null space components, showing the FDM model removes measurement noise and also synthesizes a new signal in the null space.}
    \label{fig:FDM_process}
\end{figure}
Previous research has explored various techniques for improving the resolution of PET images, including the application of super-resolution methods based on deep learning \cite{kennedy2006super, song2020super}. Diffusion models have also been applied to PET denoising \cite{cui2019pet, gong2024pet, cui2024pet}. These approaches have shown promise in enhancing image quality by reducing noise and improving clarity. However, no studies have applied generative diffusion models to the task of synthesizing higher frequency signals than what naturally occurs in the measurements. Generative diffusion models are uniquely well suited to the task of synthesizing or extrapolating the high-frequency content. The resulting images could provide clinicians with more detailed and useful information for diagnosis and treatment planning.

We propose a novel GSR-PET image processing algorithm using Fourier Diffusion Models for the time-dependent modulation transfer function (MTF) and noise power spectrum (NPS) based on our previous work in CT imaging \cite{tivnan2023fourier}. Unlike traditional diffusion models, which use white noise in both the forward and reverse processes, our method is designed to match the same MTF and NPS as the measured images at the final time step. As a result,  the trained model is a direct diffusion bridge from the distribution of low-resolution images to the distribution of high-resolution, low-noise images. Our approach is capable of generating posterior samples of high-resolution PET images from low-resolution inputs, offering a significant improvement over existing methods.


\section{METHODS}

\subsection{Fourier Diffusion Models for GSR-PET Imaging}

Our forward model for low-resolution PET images given high-resolution PET images is a linear shift-invariant system plus additive stationary Gaussian noise defined as follows:


\begin{equation}
\mathbf{y} = \mathbf{H}_\text{MTF} \mathbf{x}_0 + \boldsymbol{\Sigma}_\text{NPS}^{1/2} \boldsymbol{\epsilon}
\end{equation}


\noindent where $\mathbf{x}_0$ is a flattened vector representation of high-resolution PET image from the training data distribution, $\mathbf{x}_1$ is a vector representing the low-resolution PET image, $\boldsymbol{\epsilon}$ is zero-mean identity-covariance white Gaussian noise, $\mathbf{H}_\text{MTF}$ is an invertible circulant matrix representing the MTF of a linear shift-invariant system and $\boldsymbol{\Sigma}_\text{NPS}$ is a positive semi-definite circulant matrix representing the NPS of additive stationary Gaussian noise.  All matrix square roots in this work refer to the unique positive semi-definite matrix square root obtained by taking the square root of all eigenvalues while keeping eigenvectors the same. While we require $\mathbf{H}_\text{MTF}$ to be invertible, one can set the system response arbitrarily low, resulting in arbitrarily low SNR for those spatial frequency components. We define $\mathbf{N}_\text{Null}$ as an ideal isotropic high pass filter with cuttoff frequency set based on the full width at one tenth maximum of the MTF. 

While the high resolution PET training images have some noise and blur, our method only requires the relative MTF and NPS needed for the forward model of lower quality PET images. The GSR-PET imaging task is defined as generation of samples from the posterior distribution of high-resolution PET images, $\mathbf{x}_0$, conditional on a low-resolution input PET images, $\mathbf{x}_1$. We follow the methods in \cite{tivnan2023fourier} to implement a Fourier Diffusion Model (FDM) for GSR PET imaging. The model is defined by the forward stochastic process:

\begin{equation}
\mathbf{x}_t = \mathbf{H}_t \mathbf{x_0} + \boldsymbol{\Sigma}_t^{1/2} \frac{\mathbf{w}_t}{\sqrt{t}}  \quad , \quad 0 < t \leq 1
\label{eq:forward_process}
\end{equation}

\noindent where $\mathbf{x}_t$ is a vector representation of the forward process at time, $t$, $\mathbf{w}_t$ is a standard Wiener process in the same vector space with variance proportional to $t$, $\mathbf{H}_t$ is an invertible circulant matrix representing the MTF of the system, and $\boldsymbol{\Sigma}_t$ is a positive semi-definite circulant matrix representing the NPS.  We can design the time-dependent matrices, $\mathbf{H}_t$ and $\boldsymbol{\Sigma}_t$, to define a forward stochastic process that begins with the high-quality training data distribution, $\mathbf{x}_0$ at, $t=0$, and ends with the degraded low-resolution noisy images, $\mathbf{x}_1$ at time, $t=1$.


\begin{gather}
\mathbf{H}_t = \mathbf{I} + t ( \mathbf{H}_\text{MTF} - \mathbf{I})\\
\boldsymbol{\Sigma}_t = t \boldsymbol{\Sigma}_\text{NPS} + t \boldsymbol{\Sigma}_\text{Null}
\end{gather}

\noindent  The null space covariance $\boldsymbol{\Sigma}_\text{Null} = \sigma^2_\text{Null}$ is a positive semi-definite circulant matrix representing null space noise. This null space noise is added to the measurements before initializing the model to enable synthesis of high spatial frequencies in the diffusion process. Therefore, the formula FDM used to initialize the reverse process from the measurements is

\begin{equation}
\mathbf{x}_1 = \mathbf{y} + \boldsymbol{\Sigma}_\text{Null}^{1/2} \boldsymbol{\varepsilon}
\end{equation}

\noindent where $\boldsymbol{\varepsilon}$ is zero-mean identity-covariance Gaussian white noise. 

The forward stochastic process in \eqref{eq:forward_process} can be equivalently described by the following stochastic differential equation

\begin{equation}
\mathbf{dx}_t = \mathbf{F}_t \mathbf{x}_t dt + \mathbf{G}_t \mathbf{dw}_t
\end{equation}

\noindent where $\mathbf{F}_t$ and $\mathbf{G}_t$ are circulant matrices defined as follows:

\begin{gather}
    \mathbf{F}_t = \mathbf{H}_t^{'} \mathbf{H}_t^{-1} \\
    \mathbf{G}_t = [\boldsymbol{\Sigma}_t^{'} - \mathbf{F}_t \boldsymbol{\Sigma}_t - \boldsymbol{\Sigma}_t \mathbf{F}_t^T]^{1/2}
\end{gather}

\noindent where $\mathbf{H}_t^{'} = \frac{d}{dt}\mathbf{H}_t$ and $\boldsymbol{\Sigma}_t^{'} = \frac{d}{dt}\boldsymbol{\Sigma}_t$. In appendix D of the original article on score based generative models, the authors provide a general formula for the time reversed stochastic differential equations with matrix valued coefficients \cite{song2020score}. Applying that formula, the reverse stochastic differential equation for the FDM is given by 


\begin{equation}
\mathbf{dx}_t = \mathbf{F}_t \mathbf{x}_t dt 
 - \frac{1}{2}\mathbf{G}_t \mathbf{G}_t^T \nabla_{\mathbf{x}_t} \log p(\mathbf{x}_t)  + \mathbf{G}_t \mathbf{dw}_t
\end{equation}

FDMs are trained by optimizing a neural network based score estimator, $\mathbf{s}(\mathbf{x}_t, t)$ to approximate the score function, $\nabla_{\mathbf{x}_t} \log p(\mathbf{x}_t)$. FDMs can use the same training and sampling methods described in other works on score based generative models. The only unique aspect of FDMs is that the forward and reverse stochastic processes have spatial-frequency-dependent diffusion rates, as controlled by $\mathbf{G}_t$, and spatial-frequency-dependent signal decay rates as controlled by $\mathbf{F}_t$.  The trained model acts as a direct diffusion bridge model. That is, the reverse process can be efficiently initialized with the low-resolution PET images, rather than white noise initialization used by most other diffusion models. This allows for efficient sampling of the reverse process to sample from the posterior distribution of GSR-PET images given low-resolution PET inputs.

\subsection{Computational Implementation and Experimental Methods}
For the high resolution PET training data, we utilized the Alzheimer's Disease Neuroimaging Initiative (ADNI) dataset, specifically focusing on cases that employed the 18-FDG radiotracer and were captured using the High-Resolution Research Tomography (HRRT) scanner, known for its approximately 2~mm resolution. We extracted 5,000 axial PET images of shape $256\times256$ and split it into 4,000 slices for training and 1,000 slices for testing All images were normalized to have mean zero and standard deviation one to avoid dependence on radiation dose levels. To simulate low resolution PET measurements, the original high-resolution images were downsampled and blurred with an isotropic Gaussian kernel with full width at half maximum (FWHM) 4~mm representing the system MTF. The system NPS was defined by 4~mm FWHM isotropic Gaussian kernel minus a 16~mm FWHM isotropic Gaussian kernel to remove low frequency noise times a noise magnitude of 0.2 in normalized units. 

For the Fourier Diffusion Models (FDM), we utilized a convolutional U-net with skip connections, which was implemented in Pytorch using the Huggingface Diffusers package. The U-net architecture consisted of six depth levels with trainable down convolutions, and each level contained two convolutional layers. The number of output channels from each layer in the encoder was 32, 32, 64, 64, 128, and 128. Channel-wise cross-attention convolutional blocks were utilized at the deepest two layers to capture complex patterns in the PET images. The model was trained by sampling batches of high resolution PET images, sampling the forward process, and evaluating a score matching loss function as previously described \cite{tivnan2023fourier}. The training was conducted over 10,000 epochs, with 1,000 iterations per epoch, using a batch size of 16 and a learning rate of 0.0001. 

As a baseline for comparison we also implemented conditional diffusion models (CDM) and image-to-image Schr\"{o}dinger bridge (I2SB) using previously described methods \cite{liu2023, batzolis2021conditional}. We evaluated the performance of FDM and baseline methods by comparing the reconstructed images to the original high-resolution PET images. We computed the Peak Signal-to-Noise Ratio (PSNR) and the Structural Similarity Index Measure (SSIM). Additionally, we computed the posterior sample bias and standard deviation to assess the consistency and reliability of the FDM-generated images. In our final results we report the mean of these metrics and standard deviation over the population of 1,000 evaluation samples. We repeated these methods using 16 and 64 sampling steps to evaluate the benefit of FDM on reducing the number of time steps needed for GSR-PET synthesis.

\section{Results}

\begin{figure}[ht]
    \centering
    \includegraphics[width=0.99\linewidth]{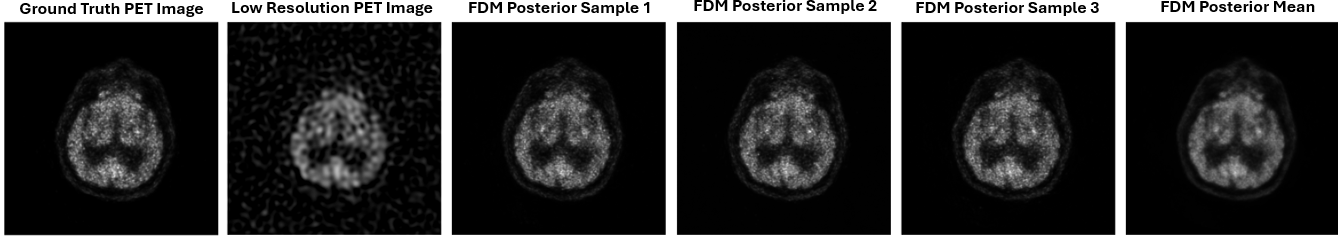}
    \caption{Multiple FDM posterior samples have the appearance of different noise realizations. The posterior mean is a high-quality PET reconstruction conditional on low-resolution measurements.}
    \label{fig:FDM_samples}
\end{figure}


\subsection{PET Image Reconstruction Results}

The PET reconstruction results demonstrate that FDMs significantly outperformed CDMs and somewhat outperformed I2SB in across all quantitative metrics as summarized in Table \ref{tab:results}. In terms of SSIM, FDMs showed superior preservation of image structure compared to both baseline methods. When considering peak signal-to-noise ratio (PSNR), FDMs provided better noise suppression and higher fidelity in the reconstructed images. The results also indicated that FDMs maintained lower posterior bias, suggesting more accurate reconstructions with minimal systematic errors. Finally, FDMs exhibited the lowest posterior standard deviation (STD), reflecting greater consistency in the reconstructions. Overall, FDMs demonstrated clear advantages over CDMs and I2SB in producing high-quality PET image reconstructions, excelling in both accuracy and consistency.

\begin{table}[ht]
\centering
\small
\caption{Quantitative Comparison of GSR PET Image Reconstruction Methods}
\label{tab:results}
\begin{tabular}{|c|c|c|c|c|c|}
\hline
\textbf{GSR Method} & \textbf{Time Steps} & \textbf{SSIM} & \textbf{PSNR} & \textbf{Posterior Bias} & \textbf{Posterior STD} \\
\hline
CDM & 16 & 0.26$\pm$0.06 & 24.2$\pm$4.3 & 0.052$\pm$0.009 & 0.029$\pm$0.04\\  \hline
I2SB & 16 & 0.63$\pm$0.10 & 27.2$\pm$5.2 & 0.022$\pm$0.004 & 0.024$\pm$0.002\\  \hline
FDM & 16 & 0.66$\pm$0.10 & 27.9$\pm$5.3 & 0.022$\pm$0.005 & 0.016$\pm$0.002\\ \hline
CDM & 64 & 0.36$\pm$0.07 & 25.7$\pm$4.6 & 0.035$\pm$0.007 & 0.023$\pm$0.003\\  \hline
I2SB & 64 & 0.67$\pm$0.11 & 27.4$\pm$5.2 & 0.022$\pm$0.004 & 0.019$\pm$0.002\\  \hline
FDM & 64 & 0.68$\pm$0.10 & 28.1$\pm$5.4 & 0.023$\pm$0.004 & 0.016$\pm$0.002\\  \hline
\end{tabular}
\end{table}


\section{CONCLUSION}

In this work, we presented a novel approach using Fourier Diffusion Models (FDMs) for Generative Super-Resolution (GSR) PET imaging. Our results demonstrated that FDMs significantly outperformed existing methods, including conditional diffusion models and image-to-image Schr\"{o}dinger bridge,  across all evaluated metrics. FDMs showed superior performance in preserving structural details, suppressing noise, and ensuring consistent and accurate reconstructions. However, our study has some limitations. We conducted our experiments using data from the High-Resolution Research Tomograph (HRRT) with a resolution of 2 mm, rather than the latest NeuroExplorer (NX-PET) data with 1 mm resolution. Additionally, our model was tested on simulated data, which may not fully capture the complexities of real-world scenarios. In ongoing work, we are applying our models to full-body PET data with approximately 4 mm resolution. We are also conducting a task-based evaluation of GSR-PET imaging for Alzheimer's Disease diagnosis, aiming to further validate the clinical applicability of our approach in more diverse and realistic settings.

\bibliography{main} 
\bibliographystyle{spiebib} 

\end{document}